\documentstyle[12pt,cite]{article}

\newcommand{\Li}{\mathop{\rm Li}\nolimits}
\newcommand{\bfe}[1]{{{\bf e}\left[#1\right]}}
\renewcommand{\Im}{\mathop{\rm Im}\nolimits}

\begin{document}

\addtolength{\baselineskip}{.3mm} \thispagestyle{empty}

\vspace{-1.5cm}
\begin{flushright}
INS-Rep-1150\\
hep-th/9607078\\
July 1996
\end{flushright}
\vspace{15mm}
\begin{center}
  {\large\sc{String Duality and Modular Forms}}\\[18mm]
  {\sc Toshiya Kawai}\\[8mm]
  {\it Institute for Nuclear Study, University of Tokyo,\\[2mm]
    Midori-cho, Tanashi,   Tokyo 188, Japan} \\[27mm]
\end{center}
\vspace{1.5cm}
\begin{center}
  {\sc Abstract}
\end{center}
\vspace{5mm}
\noindent

Tests of duality between heterotic strings on $K3\times T^2$
(restricted on certain Narain moduli subspaces) and type IIA strings
on $K3$-fibered Calabi-Yau threefolds are attempted in the weak
coupling regime on the heterotic side by identifying pertinent modular
forms related to the computations of string threshold corrections.
Concretely we discuss in parallel the three cases associated with
Calabi-Yau manifolds $(A):X(6,2,2,1,1)_{2}^{-252}$,
$(B):X(12,8,2,1,1)_{3}^{-480}$ and $(C):X(10,3,3,2,2)_{4}^{-132}$ on
the type IIA side.

\vspace{1.5cm}

\newpage

In the past year it has become harder and harder to deny that many
string theories allow dual descriptions. Through this still ongoing
development, invaluable information about non-perturbative facets of
string theory has been accumulated. String compactifications that
exhibit $N=2$ supersymmetry in four dimensions offer particularly
interesting class of examples in such string duality phenomena
\cite{rKV,rKLT,rKLM,rAGNTi,rCurio,rAFIQ,rHM,rAL,rAspinwallii, rCCLM,
  rCF,rLSTY}; heterotic strings compactified on $K3\times T^2$ may
have dual type IIA theories on Calabi-Yau manifolds that admit
structures of $K3$-fibrations \cite{rKLM}.  An extensive list of
$K3$-fibered Calabi-Yau manifolds has recently been given in
\cite{rHLY}.

In this article we shall be concerned with three (possible) $N=2$
heterotic-type IIA pairs. On the type IIA side these correspond to the
following $K3$-fibered Calabi-Yau threefolds\footnote{ Here
  $X(w_0,\ldots,w_4)_{h^{1,1}}^\chi$ denotes the Calabi-Yau manifold
  with the Hodge number $h^{1,1}$ and the Euler characteristic $\chi$
  obtained by a hypersurface of degree $\sum_i w_i$ in ${\bf
    WP}(w_0,\ldots,w_4)$.}:
\begin{center}
\begin{tabular}{ll}
$(A)$:& $X(6,2,2,1,1)_{2}^{-252}$\\[1mm]
$(B)$:& $X(12,8,2,1,1)_{3}^{-480}$\\[1mm]
$(C)$:& $X(10,3,3,2,2)_{4}^{-132}$
\end{tabular}
\end{center}
The first two cases are most familiar and candidate heterotic duals
were discovered in the pioneering work of Kachru and Vafa \cite{rKV}.
In \cite{rHM} a certain heterotic string vacuum was considered in
connection with $(B)$. This vacuum is not the one considered in
\cite{rKV} for $(B)$, thus it is not precisely the dual of type IIA
string on $(B)$.  However, in refs.\cite{rHM,rCCLM}, it was
convincingly pointed out that perturbative calculations restricted to
a particular Narain moduli {\it subspace}\ match up well with the type
IIA calculations on $(B)$.  For the third model, although we have not
yet found a precise heterotic dual in the strict sense of Kachru and
Vafa, we should like to pursue a similar story as in \cite{rHM,rCCLM}.
In the following we will describe a possible heterotic vacuum which
seems to be related to the type IIA string on $(C)$ when restricted to
a certain moduli subspace. (Thus, morally speaking, this description
should be understood as serving a motivation to write down the
expression given later.)  An ${\cal E}_8\times{\cal E}_8$ heterotic
string compactified on $K3\times T^2$ with standard embedding has
generically gauge symmetry of ${\cal E}_8\times{\cal E}_7\times
U(1)^4$ if we include graviphoton.  In total we have $248+133+4=385$
vector states.  The massless spectrum also contains 625
hypermultiplets -- 10 hypermultiplets belonging to ${\bf 56}$ of
${\cal E}_7$, the 20 $K3$ moduli hypermultiplets and the 45 gauge
bundle moduli hypermultiplets.  To give some idea of the relevance of
$(C)$, recall that the gauge symmetry ${\cal E}_7$ is attained through
enhancement of symmetry from the maximal subgroup $SO(12)\times SU(2)$
where $SO(12)$ is realized by twelve free left gauge fermions on the
world-sheet (if we adopt the fermionic formulation) and $SU(2)$ stems
from the $N=4$ superconformal algebra of the $K3$ sigma model.  We
note that ${\cal E}_7$ irreps ${\bf 133}$ and ${\bf 56}$ are
decomposed under $SO(12)\times SU(2)$ as ${\bf 133}\rightarrow ({\bf
  32_c,2})+({\bf 1,3})+({\bf 66,1})$ and ${\bf 56} \rightarrow ({\bf
  32_s,1})+({\bf 12,2})$.  Now suppose we go to the Coulomb branch of
this $SU(2)$ but retain the other non-abelian gauge symmetries ${\cal
  E}_8\times SO(12)$, thus considering a particular subspace of the
full Narain moduli space.  In this moduli subspace the number of
vector fields is $385-32\cdot 2-(3-1)=319$ and the number of
hypermultiplets is $625-10\cdot 12\cdot 2=385$. Hence twice their
difference is $2\cdot(319-385)=-132$.  On the other hand, the number
of the abelian vector fields whose scalar components consist of the
moduli fields of this Narain moduli subspace and the dilaton, is 4.
Thus this (restricted) heterotic vacuum is expected to be of relevance
to the type IIA string on a $K3$-fibered Calabi-Yau threefold with
$h^{1,1}=4$ and $\chi=-132$.

If we go to the Coulomb branch of ${\cal E}_8$ on the heterotic side
for the above three cases $(A)$, $(B)$ and $(C)$, we will get theories
with the number of abelian vector multiplets (parametrizing the
pertinent moduli subspaces) increased by eight. For these theories
possible dual type IIA theories can be identified without difficulty.
They correspond to the following $K3$-fibered Calabi-Yau
manifolds\footnote{These Calabi-Yau manifolds and their associated
  heterotic duals have already appeared in refs.\cite{rAFIQ, rHM}.}:
\begin{center}
\begin{tabular}{ll}
$(A')$:& $X(30,20,8,1,1)_{10}^{-732}$\\[1mm]
$(B')$:& $X(42,28,12,1,1)_{11}^{-960}$\\[1mm]
$(C')$:& $X(30,16,12,1,1)_{12}^{-612}$
\end{tabular}
\end{center}
For instance, in the case $(C')$ the counting on the heterotic side
goes as follows: the number of relevant abelian vector multiplets is
$4+8=12$ and the difference between the number of vector fields and
that of the hypermultiplets is $[385-(248-8)-32\cdot
2-(3-1)]-[625-10\cdot 12\cdot 2]=-306$, thus the predicted Euler
characteristic is $2\cdot (-306)=-612$.

As has been vigorously studied over the years \cite{rMirror}, for type
IIA string compactified on a (not necessarily $K3$-fibered)
Calabi-Yau manifold, the powerful techniques of mirror transformations
\cite{rCDGP}\ make it possible, if $h^{1,1}$ is sufficiently small,
the non-perturbatively exact computation of the prepotential\footnote{
  Here the $\kappa_{ijk}$ are the triple intersection numbers and we
  have omitted lower order polynomial terms.  The polylogarithm
  $\Li_k$ is defined by $\Li_k(x)=\sum_{n=1}^\infty\frac{x^n}{n^k}$.
  We will write $\bfe{x}$ for $e^{2\pi i x}$. }
\begin{equation}
  \label{IIAprep}  
{}^{\rm II}{\cal F}(t)=
\frac{1}{3 !}\sum_{i,j,k}\kappa_{ijk}t^it^jt^k+ \frac{1}{(2\pi i)^3}
   \sum_{d\in {\bf S}}N^r(d)\Li_3\left(\bfe{d\cdot t}\right)\,,
\end{equation}
as well as the topological one-loop free energy \cite{rBCOV}
\begin{eqnarray}
 \label{IIAF1}
&&{}^{\rm II}F_1^{\rm top}(t)=
-\frac{2\pi i}{12}\int c_2\wedge (t\cdot J)
+\frac{1}{6}\sum_{d\in {\bf S}} 
N^{r,e}(d)\Li_1\left(\bfe{d\cdot t}\right)\,,\\
&&N^{r,e}(d)= N^r(d)+12\sum_{d'\in {\bf S},\ d'\le d}N^e(d')\,,
\end{eqnarray}
where $t=(t^1,\ldots,t^{h^{1,1}})$ are the K{\" a}hler moduli
parameters and $J=(J_1,\ldots,J_{h^{1,1}})$ are the integral
generators of the complexified K{\" a}hler cone.  In the above we have
introduced a {\it partially ordered\/} set $({\bf S},\le)$ where ${\bf
  S}={\bf Z}_{\ge 0}^{h^{1,1}}\setminus \{ \bf 0\}$ and $d'\le d\ 
(d,d'\in {\bf S}) \Leftrightarrow {}^\exists n\in {\bf Z}_{> 0}\ {\it
  s.t.}\ d=nd'$.  The integers $N^r(d)$ and $N^e(d)$ count the virtual
numbers of rational and elliptic world-sheet instantons of multidegree
$d$.  Note that $N^e(d)=\frac{1}{12}\sum_{d'\le d} \mu(d',d)
\left[N^{r,e}(d')-N^r(d')\right]$ where $\mu(\cdot,\cdot)$ is the M{\"
  o}bius function on ${\bf S}$.

If the ${}^{\rm II}{\cal F}$ and ${}^{\rm II}F_1^{\rm top}$ of a given
$K3$-fibered Calabi-Yau manifold are exactly known and if we have
already identified a dual heterotic string theory, it is most
desirable to compare these type IIA results with heterotic
perturbation theory which is not corrected beyond one-loop. Such
comparison has been attempted in refs.\cite{rKLT,rAGNTi,rHM,rAFIQ}
with satisfactory results.

In particular, the work of Harvey and Moore \cite{rHM}\ has made it
clear that a beautiful picture emerges in perturbation theory of
heterotic strings on $K3\times T^2$: {\it the coefficients of modular
  forms appearing in the calculation of threshold corrections are
  related to rational and elliptic instanton numbers in the type IIA
  setting and in some cases are also related to the root
  multiplicities of generalized Kac-Moody (super) algebras.} Thus, it
seems that understanding this trinity of apparently remote
mathematical concepts is one of the keys to unravel the mystery of
string duality.

In this work we will attempt to pursue this line of thoughts for the
cases $(A)$, $(B)$ and $(C)$ in parallel. We will present the explicit
formulas of relevant modular forms and relate their coefficients to
perturbative heterotic prepotentials and gravitational Wilsonian
couplings which are to be compared with ${}^{\rm II}{\cal F}$ and
${}^{\rm II}F_1^{\rm top}$ in the tests of duality conjectures.
Unfortunately since the type IIA calculation of ${}^{\rm II}{\cal F}$
and ${}^{\rm II}F_1^{\rm top}$ for $(C)$ seems not to be available in
the literature (and I have not tried to work it out by myself), the
test of string duality for $(C)$ is yet to be completed.  We leave
this as a future problem.  We should also mention that part of our
results (especially for (B)) is somewhat repetitive \footnote{The case
  $(B)$ has been treated in refs.\cite{rHM,rCCLM}.}.  This is to give
a unified treatment for all the cases $(A)$, $(B)$ and $(C)$. We will
also briefly touch upon the cases $(A')$, $(B')$ and $(C')$.

Let us begin by reviewing the perturbative moduli space of heterotic
string on $K3\times T^2$ \cite{rDKLL,rAFGNT,rHM}.  The Narain lattice
$M$ of signature $(2,r)$ may be decomposed as $M=H\oplus \Lambda$.
Here $H$ is the unique even unimodular lattice of signature $(1,1)$
generated by $e_1$ and $e_2$ whose Gram matrix is given by
$\pmatrix{0&1\cr 1&0\cr}$ and $\Lambda$ is a rational\footnote{Here
  ``rational'' means that the entries of the Gram matrix of a basis
  are rational.}\ lattice of signature $(1,r-1)$ which becomes
integral after being suitably scaled.  Consider
\begin{equation}
   \label{domain}
  {\cal D}=\{ [\omega]\in {\bf P}(M\otimes {\bf C}) \mid 
  \omega^2=0\, , \quad \omega\cdot\bar\omega>0\}\,,
\end{equation}
and one of its connected component ${\cal D}^+$.  Then ${\cal
  D}^+=\Lambda\otimes {\bf R}+ iC^+(\Lambda)\simeq O(2,r)/(O(2)\times
O(r))$ where $C^+(\Lambda)$ is one of the components ({\it future
  light cone}) of $C(\Lambda):=\{x\in \Lambda\otimes {\bf R} \mid
x^2>0\}$.  The Narain moduli space is ${\cal D}^+$ divided by the
$T$-duality group (= the automorphism group of the Narain lattice).
The hermitian symmetric space ${\cal D}^+$ is also known as a bounded
domain of type IV and typically appears as the domain of the period
map of a $K3$. (See for instance \cite{rBPV,rDolgachev}.)  In this case,
the conditions $\omega^2=0$ and $\omega\cdot\bar\omega>0$ in
(\ref{domain}) are the Riemann-Hodge bilinear relations, the (suitably
scaled) $M$ is the lattice of transcendental 2-cycles and the
$T$-duality group is the monodromy group of the period map.  The
appearance of the $K3$ moduli space may be foreseen in view of
heterotic-type IIA duality since the perturbative regime of heterotic
string is, in the type IIA picture, the region where the base ${\bf
  P}^1$ of the $K3$ fibration blows up and only the fiber $K3$ becomes
relevant \cite{rAL,rAspinwallii}.  To be more precise, when we
interpret ${\cal D}^+$ as the domain of a period map, the relevant
$K3$ is a {\it mirror\/} \cite{rDolgachev}\ of the fiber $K3$ of the
$K3$-fibration in the type IIA setting. In terms of this fiber $K3$
the (suitably scaled) $\Lambda$ is the lattice of algebraic cycles,
{\it i.e.} the Picard lattice.

Notice that for $y\in {\cal D}^+$, the $\omega$ in ({\ref{domain}) is
  parametrized by
\begin{equation}
  \omega(y)=e_1-\frac{y^2}{2}e_2+y\,,
\end{equation}
since $(\omega\cdot\bar\omega)(y)=2(\Im y)^2>0$.  This parametrization
can easily be understood on the heterotic side if we recall the
classical prepotential is given by ${}^{\rm het}{\cal F}_{\rm
  cl}=\frac{1}{2}Sy^2$ where $S$ is the dilaton and take the
$T$-duality manifest basis of the period vector after a suitable
symplectic transformation \cite{rDKLL}. Then $\omega$ is the {\it
  electric\/} part of the period vector.

In the perturbative calculations on the heterotic side, say those of
threshold corrections, the following (manifestly $T$-duality
invariant) formulas of the spectrum are important:
\begin{eqnarray}
  &&p_R^2-p_L^2=\lambda^2\,,\\[2mm]
  &&\frac{1}{2}p_R^2=\frac{\vert\lambda\cdot\omega\vert^2}
     {\omega\cdot\bar\omega}\,,\quad \omega=\omega(y)\,,
\end{eqnarray}
where $(p_R,p_L)$ are the right-left momenta of the compactified
sector and $\lambda\in M$ and $y \in {\cal D}^+$. The second formula
gives the mass formula of BPS saturated string elementary states and
$\lambda\cdot\omega$ is the central charge appearing in 4D $N=2$
superalgebra.  If the central charge vanishes at some point in the
moduli space, extra massless BPS states appear there in general. In
the heterotic picture this occurs where symmetry enhancement arises
through the Frenkel-Kac construction. In the type II picture
$\lambda\cdot\omega$ vanishes if some 2-cycles of the $K3$ collapse
and the $K3$ develops ${\cal ADE}$ singularities.  ({\it cf.}
\cite{rWitten}.)

Now we turn to the specific cases of $(A)$, $(B)$ and $(C)$.  For
these cases the lattice $\Lambda$ is given respectively by
\begin{eqnarray}
  &&\Lambda_A=L_+\\
  &&\Lambda_B=H'\\
  &&\Lambda_C=H'\oplus L_-
\end{eqnarray}
where $H'$ is a copy of $H$ and its basis is denoted by $\{f_1,f_2\}$
and $L_\pm$ are the one-dimensional lattices generated by $\delta_\pm$
with $\delta_\pm$ satisfying $\delta_\pm^2=\pm\frac{1}{2}$.  For cases
$(A')$, $(B')$ and $(C')$ we should make replacement: $\Lambda_*
\rightarrow \Lambda_*\oplus {\cal E}_8(-1)$ where ${\cal E}_8(-1)$ is
the negative of an ${\cal E}_8$ root lattice.

Since $r=1$, $2$, and $3$ for $(A)$, $(B)$ and $(C)$, the space ${\cal
  D}^+$ is given respectively by ${\bf H}_1$, ${\bf H}_1\times {\bf
  H}_1$ and ${\bf H}_2$ where ${\bf H}_1$ is the standard upper
half-plane and ${\bf H}_2$ is the Siegel upper half-plane of genus
two.  The $T$-duality groups are respectively $SL(2,{\bf Z})$,
$(SL(2,{\bf Z}) \times SL(2,{\bf Z}))/{\bf Z}_2$ and $Sp(4,{\bf Z})$.
We will take the following parametrization of $y\in {\cal D}^+$:
\begin{eqnarray}
  &(A):&  y=2T\delta_+\, ,\quad T\in {\bf H}_1\,,\\
  &(B):&  y=Tf_1+Uf_2\, ,\quad (T,U)\in {\bf H}_1\times {\bf H}_1\,,\\
  &(C):&  y=Tf_1+Uf_2+2V\delta_- \, ,\quad \Omega=
   \pmatrix{T&V\cr V&U\cr}\in {\bf H}_2\,.
\end{eqnarray}

To present our formulas of modular forms we have to introduce some
notations.  First recall that the Jacobi theta functions are defined
by
\begin{equation}
  \begin{array}{lll}
&\displaystyle \vartheta_1(\tau,z)=
i \sum_{n\in {\bf Z}}
(-1)^n q^{\frac{1}{2}(n-\frac{1}{2})^2} \zeta^{n-\frac{1}{2}}\,,\quad
&\displaystyle \vartheta_2(\tau,z)=
\sum_{n\in {\bf Z}}q^{\frac{1}{2}(n+\frac{1}{2})^2} 
\zeta^{n+\frac{1}{2}}\,,\\[1mm]
&\displaystyle \vartheta_3(\tau,z)=
\sum_{n\in {\bf Z}}q^{\frac{n^2}{2}} \zeta^n\,, \quad 
&\displaystyle \vartheta_4(\tau,z)=
\sum_{n\in {\bf Z}}(-1)^n q^{\frac{n^2}{2}} \zeta^n\,, 
  \end{array}
\end{equation}
where $q=\bfe{\tau}$ and $\zeta=\bfe{z}$.  For later convenience we
also introduce the following theta functions:
\begin{equation}
\theta_{\rm ev}(\tau,z)=
\sum_{n\in {\bf Z}}q^{\frac{1}{4}(2 n)^2} \zeta^{2 n}\,,\qquad
\theta_{\rm od}(\tau,z)=
\sum_{n\in {\bf Z}}q^{\frac{1}{4}(2 n+1)^2} \zeta^{2 n+1}
\end{equation}
thus $\theta_{\rm ev}(\tau,z)=\vartheta_3(2\tau,2z)$ and
$\theta_{\rm od}(\tau,z)=\vartheta_2(2\tau,2z)$.  We use simplified
notations: $ \vartheta_k^0(\tau)=\vartheta_k(\tau,0)$,
$\theta_{\rm ev}^0(\tau) =\theta_{\rm ev}(\tau,0)$ and $
\theta_{\rm od}^0(\tau)=\theta_{\rm od}(\tau,0)$.

As is well-known the ring of modular forms with respect to $SL(2,{\bf
  Z})$ is generated by the Eisenstein series of weight four and six:
\begin{equation}
  E_4(\tau)=1+240\sum_{n=1}^\infty\sigma_3(n)q^n\,,\quad 
  E_6(\tau)=1-504\sum_{n=1}^\infty\sigma_5(n)q^n\, ,
\end{equation}
where $\sigma_k(n)=\sum_{d\mid n}d^k$.  In addition, we need the
Eisenstein series of ``weight two'':
\begin{equation}
  E_2(\tau)=1-24\sum_{n=1}^\infty\sigma_1(n)q^n=
\Theta_q\log\Delta(\tau)\,,
\end{equation}
where $\Delta(\tau)=\eta(\tau)^{24}$ and $\Theta_q$ is the Euler
derivative $q\frac{\displaystyle d}{\displaystyle dq}$.  It satisfies
the functional equation $E_2\left(\frac{a\tau+b}{c\tau+d}\right)=
(c\tau+d)^2E_2(\tau)+\frac{12}{2\pi i}c(c\tau+d)$ for ${
  \pmatrix{a&b\cr c&d\cr}}\in SL(2,{\bf Z})$.  These Eisenstein series
are mutually related by
\begin{equation}
  \Theta_q E_k=\frac{k}{12}(E_2E_k-E_{k+2})\,,\quad (k=4,6)\,,
\end{equation}
where $E_8=E_4^2$.  The elliptic modular
function is given by
\begin{equation}
  j(\tau)=\frac{E_4(\tau)^3}{\Delta(\tau)}=
   \frac{E_6(\tau)^2}{\Delta(\tau)}+j(i)\,, \quad j(i)=1728\, .
\end{equation}

For our purpose, it is useful to introduce the following functions
({\it cf.} \cite{rBorcherds})
\begin{eqnarray}
  &&\theta(\tau)=\sum_{n\in {\bf Z}}q^{n^2/4}
=\vartheta_3^0\left(\tau/2\right)
=\theta_{\rm ev}^0(\tau)+\theta_{\rm od}^0(\tau)\,,\\
  &&F(\tau)=\sum_{n>0,\,n\, {\rm odd}}\sigma_1(n)q^{n/4}
=\frac{1}{16} \vartheta_2^0\left(\tau/2\right)^4\,.
\end{eqnarray}
They satisfy the functional equations
\begin{eqnarray}
  &&\theta(\tau+4)
=\theta(\tau)\,,\quad \theta\left(\frac{\tau}{\tau+1}\right)=
(\tau+1)^{\frac{1}{2}}\,\theta(\tau)\,,\\
  &&F(\tau+4)=F(\tau)\,,\quad F\left(\frac{\tau}{\tau+1}\right)=
(\tau+1)^2\, F(\tau)\,,
\end{eqnarray}
and hence are modular forms with respect to the modular subgroup
$\Gamma^0(4)=\Bigl\{{\pmatrix{a&b\cr c&d\cr}} \in SL(2,{\bf Z}) \Bigm|
b \equiv 0 \bmod{4}\Bigr\}$.

In what follows we will give our expressions of modular forms for
cases $(A)$, $(B)$ and $(C)$. There are three kinds of (nearly
holomorphic) modular forms $H_*$, $\tilde H_*$ and $J_*$ for $*=A$,
$B$, $C$.  The constant terms of $H_*$, $\tilde H_*$ and $J_*$ are
respectively $\chi$, $\chi-48$ and $0$ where $\chi$ is the Euler
characteristic of the corresponding Calabi-Yau manifold and we remind
that $b_{\rm grav}=48-\chi$ is the gravitational one-loop beta
coefficient \cite{rAGN, rAGNT}.  The functions $H_*$, $\tilde H_*$ and
$J_*$ are related in such a way that will turn out to be important.

For case $(A)$ our proposed expressions are
\begin{eqnarray}
 &&H_A(\tau)=\frac{2\theta(\tau)E_4(\tau)G_6(\tau)}{\Delta(\tau)}=
\sum_{N\in{\bf Z}\ {\rm or}\ {\bf Z}+\frac{1}{4}}c(N)q^{N}\\
&&\ = {2\over q} - 252 - 2496\,{q^{{1/4}}} - 223752\,q - 
  725504\,{q^{{5/4}}} -\cdots\\
&&\tilde H_A(\tau)=
\frac{2\theta(\tau)E_2(\tau)E_4(\tau)G_6(\tau)}{\Delta(\tau)}=
\sum_{N\in{\bf Z}\ {\rm or}\ {\bf Z}+\frac{1}{4}}{\tilde c}(N)q^{N}\\
&&\ =\frac {2}{{q}} - 300 - 2496\,{q}^{1/4} - 
217848\,{q} - 665600\,{q}^{5/4}- \cdots \\
&& J_A(\tau)
=2\theta(\tau)\left(\frac{E_6(\tau)G_6(\tau)}{\Delta(\tau)}+
870\right)
=\sum_{N\in{\bf Z}\ {\rm or}\ {\bf Z}+\frac{1}{4}}a(N)q^{N}\\
&&\ =\frac {2}{{q}} + 984\,{q}^{1/4} + 
286752\,{q} + 1131520\,{q}^{5/4} +\cdots
\end{eqnarray}
where
\begin{equation}
  G_6(\tau)=E_6(\tau)
-2 F(\tau) (\theta(\tau)^4-2 F(\tau)) (\theta(\tau)^4-16 F(\tau))\,,
\end{equation}
and we have the relation
\begin{equation}
\label{relA}
 -\frac{24}{3} \Theta_q\, H_A(\tau)
=\tilde H_A(\tau)+7\, J_A(\tau)+300\,\theta(\tau)\,.
\end{equation}

Similarly for $(B)$ we have \cite{rHM}
\begin{eqnarray}
  &&H_B(\tau)
=\frac{2E_4(\tau)E_6(\tau)}{\Delta(\tau)}=\sum_{N\in {\bf Z}}c(N)q^N\\
  &&\ =\frac{2}{q} - 480 - 282888\,{q} - 17058560\,{q}^{2}-\cdots\\
  &&\tilde H_B(\tau)
=\frac{2E_2(\tau)E_4(\tau)E_6(\tau)}{\Delta(\tau)}
=\sum_{N\in {\bf Z}}\tilde c(N)q^N\\
  &&\ =\frac{2}{q} - 528 - 271512\,{q} - 10234880\,{q}^{2} -\cdots\\
  &&J_B(\tau)=2\left(\frac{E_6(\tau)^2}{\Delta(\tau)}+984\right)
=\sum_{N\in {\bf Z}}a(N)q^N\\
&&\ =\frac{2}{q}+ 393768\,{q} + 42987520\,{q}^{2}+\cdots
\end{eqnarray}
and 
\begin{equation}
\label{relB}
  -\frac{24}{4} \Theta_q\, H_B(\tau)
=\tilde H_B(\tau)+5\, J_B(\tau)+528\,.
\end{equation}

To present our expressions for $(C)$, some familiarity with Jacobi
forms \cite{rEZ}\ is needed.  A Jacobi form $\Phi_{k,m}$ of weight $k$
and index $m$ satisfies
\begin{eqnarray}
  &&\Phi_{k,m}\left(\frac{a\tau+b}{c\tau+d},\frac{z}{c\tau+d}\right)=
 (c\tau+d)^k\bfe{\frac{m c z^2}{c\tau+d}}\Phi_{k,m}(\tau,z)\,, \\
  &&\Phi_{k,m}(\tau,z+\lambda\tau+\mu)
=\bfe{-m(\lambda^2\tau+2\lambda z)}
\Phi_{k,m}(\tau,z)\,,
\end{eqnarray}
where ${\pmatrix{a&b\cr c&d\cr}}\in SL(2,{\bf Z})$ and $\lambda,\,
\mu\in {\bf Z}$.  The ring of Jacobi forms of index $1$ is generated
by the Jacobi-Eisenstein series $E_{4,1}$ of weight 4 and $E_{6,1}$ of
weight 6 which have expansions
\begin{eqnarray}
  &&E_{4,1}(\tau,z)=1 +  \left(\frac {1}{{\zeta}
^{2}}  + \frac {56}{{\zeta}}+ 126 + 56{\zeta}+{\zeta}^{2}\right){q} 
+\cdots\\
  &&E_{6,1}(\tau,z)=1 +  \left(\frac {1}{{\zeta}
^{2}} -  \frac {88}{{\zeta}}- 330 - 88{\zeta}+{\zeta}^{2} \right){q}
 +\cdots\,.
\end{eqnarray}
The $K3$ elliptic genus $Z(\tau,z)$ is a (weak) cusp Jacobi form of
weight $0$ and index $1$ given by
\begin{equation}
  \begin{array}{ll}
&\displaystyle Z(\tau,z)= 
 \frac{1}{72}\frac{E_4(\tau)^2E_{4,1}(\tau,z)-E_6(\tau)E_{6,1}(\tau,z)}
{\Delta(\tau)}\\
& \displaystyle=\frac{2}{\zeta}+20+2 \zeta
+\left(\frac{20}{\zeta^2}-\frac{128}{\zeta}+216
-128\zeta+20\zeta^2\right)\, q+\cdots
  \end{array}
\end{equation}

A general theory shows that any Jacobi form $\Phi_{k,1}$ of index $1$
can be decomposed as
\begin{equation}
  \Phi_{k,1}(\tau,z)=\Phi_{k,1}^{\rm ev}(\tau)\theta_{\rm ev}(\tau,z)+
  \Phi_{k,1}^{\rm od}(\tau)\theta_{\rm od}(\tau,z)\,,
\end{equation}
and we introduce 
\begin{equation}
  \hat\Phi_{k,1}(\tau):=\Phi_{k,1}^{\rm ev}(\tau)+\Phi_{k,1}^{\rm od}(\tau)\,,
\end{equation}
for such a decomposition.  Explicitly we have \cite{rKYY}
\begin{eqnarray}
  && Z(\tau,z)=Z^{\rm ev}(\tau)\theta_{\rm ev}(\tau,z)
       +Z^{\rm od}(\tau)\theta_{\rm od}(\tau,z)\\[1mm]
  &&Z^{\rm ev}(\tau)
   =\frac{6\{\vartheta_2^0(\tau)\vartheta_4^0(\tau)\}^2
\theta_{\rm ev}^0(\tau)
   -2(\vartheta_4^0(\tau)^4-\vartheta_2^0(\tau)^4)\theta_{\rm od}^0(\tau)}
    {\eta(\tau)^6} \nonumber \\[1mm]
  &&\quad = 20 + 216\,{q} + 1616\,{q}^{2} + 8032\,{q}^{3} 
+\cdots \\[1mm]
  &&Z^{\rm od}(\tau)
   =\frac{6\{\vartheta_2^0(\tau)\vartheta_4^0(\tau)\}^2
\theta_{\rm od}^0(\tau)
   +2(\vartheta_4^0(\tau)^4-\vartheta_2^0(\tau)^4)\theta_{\rm ev}^0(\tau)}
   {\eta(\tau)^6} \nonumber \\[1mm]
  &&\quad =  \frac {2}{{q}^{1/4}} - 128\,{q}^{3/4} - 1026\,
           {q}^{7/4} - 5504\,{q}^{11/4} -\cdots\,,
\end{eqnarray}
and
\begin{eqnarray}
  &&E_{4,1}(\tau,z)=E_{4,1}^{\rm ev}(\tau)\, \theta_{\rm ev}(\tau,z)
                  +E_{4,1}^{\rm od}(\tau)\, \theta_{\rm od}(\tau,z)\\
  &&E_{4,1}^{\rm ev}(\tau)=
\theta_{\rm ev}^0(\tau)^7+7 \theta_{\rm ev}^0(\tau)^3\theta_{\rm od}^0(\tau)^4 
\nonumber\\
&&\quad\quad=1 + 126\,{q} + 756\,{q}^{2} +\cdots\\
 &&E_{4,1}^{\rm od}(\tau)=
\theta_{\rm od}^0(\tau)^7+7 \theta_{\rm od}^0(\tau)^3\theta_{\rm ev}^0(\tau)^4 
\nonumber\\
&&\quad\quad=56\,{q}^{3/4} + 576\,{q}^{7/4} + 1512\,{q}^{11/4} 
+\cdots\\
 &&E_{6,1}(\tau,z)=E_{6,1}^{\rm ev}(\tau)\, \theta_{\rm ev}(\tau,z)
                  +E_{6,1}^{\rm od}(\tau)\, \theta_{\rm od}(\tau,z)\\
 &&E_{6,1}^{\rm ev}(\tau)=-\frac{1}{4}
\left[\vartheta_2^0(\tau)^6 Z^{\rm ev}(\tau)
 -(\vartheta_3^0(\tau)^6+\vartheta_4^0(\tau)^6)Z^{\rm od}(\tau)\right]
\eta(\tau)^6\nonumber\\
 &&\qquad=1 - 330\,{q} - 7524\,{q}^{2} -\cdots\\
 &&E_{6,1}^{\rm od}(\tau)=-\frac{1}{4}
\left[-\vartheta_2^0(\tau)^6 Z^{\rm od}(\tau)
 +(\vartheta_3^0(\tau)^6-\vartheta_4^0(\tau)^6)Z^{\rm ev}(\tau)\right]
\eta(\tau)^6  \nonumber\\
 &&\qquad= - 88\,{q}^{3/4} - 4224\,{q}^{7/4} - 30600\,{q}^{11/4} 
-\cdots
\end{eqnarray}

With this preparation we can write down the expressions for $(C)$:
\begin{eqnarray}
  &&H_C(\tau)=\frac{2E_4(\tau)\hat E_{6,1}(\tau)}{\Delta(\tau)}
=\sum_{N \in\, {\bf Z}\ {\rm or}\ {\bf Z}+\frac{3}{4}} c(N)q^{N}\\
&&\ = \frac {2}{{q}} - \frac {176}{{q}^{1/4}} 
- 132 - 54912\,{q}^{3/4} - 172800\,{q} - 
3742416\,{q}^{7/4} - \cdots\\
  &&\tilde H_C(\tau)
=\frac{2E_2(\tau)E_4(\tau)\hat E_{6,1}(\tau)}{\Delta(\tau)}
=\sum_{N\in\, {\bf Z}\ {\rm or}\ {\bf Z}
+\frac{3}{4}}\tilde c(N)q^{N}\\
  &&\ =\frac {2}{{q}}
 - \frac {176}{{q}^{1/4}} - 180 - 50688\,{q}^{3/4} - 169776\,{q} - 
2411856\,{q}^{7/4}-\cdots\\
  &&J_C(\tau)=\frac{2E_6(\tau)\hat E_{6,1}(\tau)}{\Delta(\tau)}+
 81\hat Z(\tau)=\sum_{N\in\, {\bf Z}\ {\rm or}\ {\bf Z}
+\frac{3}{4}}a(N)q^{N}\\
  &&\ =\frac {2}{{q}} - \frac {14}{{q}^{1/4}} + 65664\,{q}^{3/4} 
+ 262440\,{q} + 8909838\,{q}^{7/4} +\cdots
\end{eqnarray}
Again there exists a relation among these functions, {\it i.e.}
\begin{equation}
   \label{relC}
  -\frac{24}{5}\Theta_q\,H_C(\tau)=
  \tilde H_C(\tau)+\frac{19}{5}J_C(\tau)+9\hat Z(\tau)\,.
\end{equation}
This  follows from
\begin{equation}
    (\Theta_q-\frac{1}{4}\Theta_\zeta^2)\,E_{k,1}
=\frac{2k-1}{24}
    (E_2 E_{k,1}-E_{k+2,1})\,,\quad (k=4,6)\,,  
\end{equation}
where $E_{8,1}=E_4 E_{4,1}$ and 
\begin{equation}
  (\Theta_q-\frac{1}{4}\Theta_\zeta^2)\,\theta_{\rm ev}
=(\Theta_q-\frac{1}{4}\Theta_\zeta^2)\,\theta_{\rm od}=0\,.
\end{equation}

Having presented our expressions for modular forms, we can now discuss
the physical implications of the coefficients of these modular forms.
The heterotic prepotential ${}^{\rm het} {\cal F}$ assumes the form
\begin{equation}
\label{het-prep}
 {}^{\rm het} {\cal F}(S,y)
=\frac{1}{2}Sy^2+v(y)+{\cal F}_{NP}(\bfe{S},y)\,.
\end{equation}
As in \cite{rHM}, the perturbative contribution $v(y)$ should be
written in terms of the coefficients of $H_*$ as
\begin{equation}
  \label{het-pert-prep}
  v(y)=p(y)-\frac{1}{(2\pi i)^3}\sum_{\alpha\in \Lambda,\, \alpha>0}
   c(\alpha^2/2)\,\Li_3(\bfe{\alpha\cdot y})\,,
\end{equation}
where  $\alpha>0$ means that 
\begin{equation}
  \begin{array}{ll}
(A):& n>0, \\
(B):& (i)\ k>0,\ {\rm or}\ (ii)\ k=0,\ l>0, \\
(C):& (i)\ k>0,\ {\rm or}\ 
       (ii)\ k=0,\ l>0,\ {\rm or}\ 
       (iii)\ k=l=0,\ b<0\,,
  \end{array}
\end{equation}
if $\alpha$ is parametrized as $(A):$ $\alpha=n\delta_+$, $(B):$
$\alpha=lf_1+kf_2$, $(C):$ $\alpha=lf_1+kf_2-b\delta_-$.  The term
$p(y)$ is a chamber-dependent \cite{rHM}\ cubic polynomial and for
each case we can take\footnote{$p_B(y)$ was
  calculated in \cite{rHM}.  A similar calculation leads to $p_C(y)$.}
\begin{eqnarray}
  &&p_A(y)=\frac{2}{3}T^3-T^2-\frac{13}{6}T\,,\\
  &&p_B(y)={ \frac {1}{3}}{U}^{3} - 
{ \frac {11}{6}}{U}^{2}{T} - {U}{T}^{2}\,,\\
  &&p_C(y)= p_B(y) - \frac {31}{6}{U}{V}^{2} - 5{T}{V}^{2} 
+  \frac {43}{6}{T}{U}{V}+\frac {37}{6}{V}^{3}\,,
\end{eqnarray}
where $p_B$ and $p_C$ are evaluated in a chamber where $\Im T>\Im U$.
However $p(y)$ is ambiguous due to the freedom of adding quadratic
polynomials in the components of $\omega(y)$. Thus, for instance, we
have
\begin{equation}
  p_C(y)\sim \frac {1}{3}{U}^{3}-6TV^2-7UV^2+\frac{40}{3}V^3\,.
\end{equation}

Next we turn to ${}^{\rm het}{\cal F}_1$ which is the heterotic
equivalent of ${}^{\rm II}F_1^{\rm top}$.  For this purpose we need
several product formulas of automorphic forms on $T$-duality groups.
Such product representations have recently been the subject of
intensive study by several mathematicians
\cite{rBorcherds,rGNi,rGNii,rGNiii}\ and possible connections with the
denominator functions of generalized Kac-Moody (super) algebras have
been discussed.  For a given series $\varphi(\tau)=\sum_Nc(N)q^N$
introduce an infinite product $\Psi$ by
\begin{equation}
  \Psi[\Lambda,\rho,\varphi]
=\bfe{\rho\cdot y}\prod_{\alpha\in\Lambda,\,
   \alpha>0}(1-\bfe{\alpha\cdot y})^{c(\alpha^2/2)}\,,
\end{equation}
then what is relevant to us may be summarized as:
\begin{equation}
\label{prodformula}
  \begin{array}{lccc}
 \Lambda   &  \varphi     &\  \rho                 & \Psi         
  \\[3mm] 
 \Lambda_A & J_A(\tau)    &\  -\delta_+            & j(T)-j(i)     
 \\[2mm]
 \Lambda_A & \theta(\tau) &\  \frac{1}{12}\delta_+ & \eta(T)^2     
 \\[2mm]
 \Lambda_B & J_B(\tau)    &\  -2f_2                & (j(T)-j(U))^2 
 \\[2mm]
 \Lambda_B & 1            &\ \frac{1}{24}(f_1+f_2) & \eta(T)\eta(U)
 \\[2mm]
 \Lambda_C & J_C(\tau)    &\  -f_1-3f_2+5\delta_-                     
  &\displaystyle \frac{\Delta_{35}(\Omega)^2}{\Delta_5(\Omega)^{14}}
\\[2mm]
 \Lambda_C & \hat Z(\tau) &\  f_1+f_2-\delta_-     & \Delta_5(\Omega)^2
  \end{array}
\end{equation}
where we assumed $\Im T>\Im U$. The functions $\Delta_{35}(\Omega)$
and $\Delta_5(\Omega)$ are related to the Igusa cusp forms
\cite{rIgusa}, $\chi_{35}(\Omega)$ and $\chi_{10}(\Omega)$ by the
relations $\Delta_{35}(\Omega)=4i\chi_{35}(\Omega)$ and
$\Delta_5(\Omega)^2=-4\chi_{10}(\Omega)$.  The first and third results
in (\ref{prodformula}) are due to Borcherds \cite{rBorcherds}, while
the last one is due to Gritsenko and Nikulin \cite{rGNi,rGNii}.  The
 Jacobi form
\begin{eqnarray}
 &&X(\tau,z)=\frac{E_6(\tau)E_{6,1}(\tau,z)}{\Delta(\tau)}+44Z(\tau,z)
 \nonumber\\
 &&\ =\frac{1}{q} +  \left(\frac {1}{{\zeta}^{2}}+ 70 + {\zeta}^{2} 
\right)\\
&&\ \qquad +  \left( \frac {70}{{\zeta}^{2}} 
+ \frac {32384}{{\zeta}}+131976  + 32384\,{\zeta}
+ 70\,{\zeta}^{2} \right) {q} 
+\cdots \nonumber
\end{eqnarray}
coincides with the last equation in ref.\cite{rGNiii}, namely,
$\phi_{0,1}\vert_0 T_0(2)-2\phi_{0,1}$ in the notation there.
Consequently, the fifth result also follows from their result. The
correspondence $Z(\tau,z)\leftrightarrow \Delta_5(\Omega)^2$ can be
confirmed by a calculation of threshold correction \cite{rKawai}.
Similarly I have checked the correspondence $X(\tau,z)\leftrightarrow
\Delta_{35}(\Omega)$ by an evaluation of the pertinent modular
integral following the approach of \cite{rDKL}\cite{rHM}.

If we separate the heterotic free energy ${}^{\rm het}{\cal F}_1$ into
the perturbative and non-perturbative parts as
\begin{equation}
  \label{het-F1}
  {}^{\rm het}{\cal F}_1(S,y)=-\frac{2\pi i }{12} f^W(S,y)+ 
{\cal F}_1^{\rm NP}(\bfe{S},y)\,,
\end{equation}
then one may infer that
\begin{eqnarray}
  \label{grwilsonA}
&&  f_A^W(S,y)
=24\tilde S 
+\frac{2}{2\pi i}\left[7\log(j(T)-j(i))+300\log\eta(T)^2\right]\,,\\
\label{grwilsonB}
&&  f_B^W(S,y)
=24\tilde S +\frac{2}{2\pi i}\left[5\log(j(T)-j(U))^2
+528\log(\eta(T)\eta(U))\right]\,,\\
 \label{grwilsonC}
&&  f_C^W(S,y)=24\tilde S 
+\frac{2}{2\pi i}\left[ 
\frac{19}{5}\log(\Delta_{35}(\Omega)^2/\Delta_5(\Omega)^{14})
+9\log \Delta_5(\Omega)^2\right]\,.
\end{eqnarray}
In these equations,
\begin{equation}
  \tilde S=S+\frac{1}{r+2}\nabla_y^2v(y)\,,
\end{equation}
is the invariant dilaton \cite{rDKLL}\cite{rHM}, where $\nabla_y^2$ is
a second-order differential operator satisfying $\nabla_y^2\, \bfe{\alpha
  \cdot y}=(2\pi i)^2\, \alpha^2 \,\bfe{\alpha \cdot y}$ and is
explicitly given by $(A):$ $\nabla_y^2=\frac{1}{2}\partial_T^2$,
$(B):$ $\nabla_y^2=2\partial_T\partial_U$ and $(C):$
$\nabla_y^2=2(\partial_T\partial_U-\frac{1}{4}\partial_V^2)$.  One can
easily see that the expressions (\ref{grwilsonA})--(\ref{grwilsonC})
for $f^W$ have physically acceptable modular properties with respect
to $T$-duality as Wilsonian gravitational couplings.  Actually the
expressions for $f_A^W$ and $f_B^W$ can be seen to agree with the
results in \cite{rKLT, rKLM}.  Using the relations (\ref{relA}),
(\ref{relB}), (\ref{relC}) and the product formulas
(\ref{prodformula}) one can deduce for all the cases we are
considering that
\begin{equation}
 \label{grwilson-gen}
 f^W(S,y)=24 S + \ell(y) +\frac{2}{2\pi i}
\sum_{\alpha\in \Lambda,\, \alpha>0}
\tilde   c(\alpha^2/2)\,\Li_1(\bfe{\alpha\cdot y})\,, 
\end{equation}
where $\ell(y)$ is linear in $y$ and has an ambiguity due to that of
$p(y)$.

In order to test the duality conjectures we must compare
(\ref{het-prep}) and (\ref{het-pert-prep}) against (\ref{IIAprep}) as
well as (\ref{het-F1}) and (\ref{grwilson-gen}) against (\ref{IIAF1})
by judiciously identifying linear combinations of $t^i$'s with $S$ and
$y$.  For case $(B)$ this has already been done in \cite{rHM,rCCLM}.
For $(A)$\footnote{We should emphasize that the first quantitative
  test of duality conjecture for $(A)$ \cite{rKV}\ was done in
  \cite{rKLT}.}, the comparison of (\ref{het-F1}) and
(\ref{grwilson-gen}) with (\ref{IIAF1}) leads to\footnote{ For
  notational simplicity we write $N^*(n,m)$ instead of $N^*((n,m))$
  for $(n,m)\in {\bf S}$.}
\begin{equation}
 N^{r,e}(n,0)=N^r(n,0)+12 \sum_{d\mid n} N^e(d,0)=-\tilde c(n^2/4)\,,
\quad (n \geq 1)\,, 
\end{equation}
where our choice of the identification rule is such that $t^1=T$ and
$t^2=S$.  Thus we obtain conjectured relations for $(A)$:
\begin{eqnarray}
  && N^r(n,0)=-c(n^2/4)\,,\\
  && N^e(n,0)=\frac{1}{12}\sum_{d \mid n}\mu\left(\frac{n}{d}\right)\, 
[\,c(d^2/4)-\tilde c(d^2/4)\, ]\,,
\end{eqnarray}
where $n \geq 1$ and $\mu(\cdot)$ is the classical M{\" o}bius function.
These give
\[
{\begin{array}{r@{\ \ \ \ \ \ \ \ \ }r@{\ \ \ \ }r}
n & N^r(n,0) & N^e(n,0)\\[2mm]
1 & 2496 & 0 \\
2 & 223752 & -492 \\
3 & 38637504 & -1465984 \\
4 & 9100224984 & -1042943028 \\
5 & 2557481027520 & -595277880960 \\
6 & 805628041231176 & -316194811079664 \\
7 & 274856132550917568 & -163214406650542848 \\
8 & 99463554195314072664 & -83229690442895106144 
\end{array}}
\]
in perfect agreement with the type IIA results obtained in
\cite{rCOFKM,rHKTY,rHeM}.

As for case $(C)$, before attempting any comparison toward
establishment of the conjecture, we need results from type IIA
calculations for this case, which, as stated earlier, are yet to be
done.  But we wish to remark one more point about the case $(C)$. Our
proposed formulas of modular forms are related to the calculation of
threshold corrections on the heterotic side.  Analogously to the case
$(B)$ treated in \cite{rHM}, one may consider the threshold
corrections to the gauge couplings for ${\cal E}_8$ and $SO(12)$.  The
one-loop beta function coefficients for these gauge groups must appear
as the constant terms of
\begin{eqnarray}
  &&-\frac{1}{12}\left(\frac{E_2(\tau)E_4(\tau)\hat E_{6,1}(\tau)
-E_6(\tau)\hat E_{6,1}(\tau)}{\Delta(\tau)}\right)\nonumber \\
  &&\ =  - 60 + 5280\,{q}^{3/4} + 17280\,{q}+\cdots\,,\\
\noalign{\noindent and}
  &&-\frac{1}{12}\left(\frac{E_2(\tau)E_4(\tau)\hat E_{6,1}(\tau)
-E_4(\tau)^2\hat E_{4,1}(\tau)}{\Delta(\tau)}\right)  \nonumber \\
  &&\ = 12\,{\displaystyle \frac {1}{{q}^{1/4}}} + 60 + 4512\,{q}^{3/4}
+\cdots\,.
\end{eqnarray}
This is actually the case since $b_{{\cal E}_8}=-{\cal I}({\bf
  248})=-60$ and $b_{SO(12)}=10 {\cal I}({\bf 32_s})-{\cal I}({\bf
  66})=10\cdot 8 -20=60$ where ${\cal I}({\bf rep})$ denotes the index
of an irrep ${\bf rep}$.

Though we have restricted ourselves to specific cases in this work, it
may well be true that formulas such as (\ref{het-pert-prep}) and
(\ref{grwilson-gen}) have universal meanings in perturbation theory of
heterotic strings on $K3\times T^2$ as already advocated for the case
of the prepotential in \cite{rHM}. Also there seems to be still much
room for clarifying possible roles of generalized Kac-Moody (super)
algebras in a more profound understanding of string duality.

Finally we comment on the cases $(A')$, $(B')$ and $(C')$.  The
modular forms $H_{A'}$, $H_{B'}$ and $H_{C'}$ are obtained by dropping
$E_4$ from $H_{A}$, $H_{B}$ and $H_{C}$ since $E_4$ is the theta
series for the ${\cal E}_8$ root lattice\footnote{ The case $(B')$ was
  already given in \cite{rHM}.}:
\begin{eqnarray}
  &&H_{A'}(\tau,z)=\frac{2\theta(\tau)G_6(\tau)}{\Delta(\tau)}\\
  &&\quad=\frac {2}{{q}} - 732 - 2496\,{q}^{1/4} - 52392\,{q} 
- 126464\,{q}^{5/4} -\cdots\\
  &&H_{B'}(\tau,z)=\frac{2E_6(\tau)}{\Delta(\tau)}\\
  &&\quad=\frac{2}{q} - 960 - 56808\,{q} - 1364480\,{q}^{2}-\cdots\\
  &&H_{C'}(\tau,z)=\frac{2\hat E_6(\tau)}{\Delta(\tau)}\\
&&\quad=\frac {2}{{q}} -  
\frac {176}{{q}^{1/4}} - 612 - 12672\,{q}^{3/4} - 30240\,{q} - 
320976\,{q}^{7/4} -\cdots
\end{eqnarray}
The constant terms correctly reproduce the Euler characteristics of
the corresponding Calabi-Yau manifolds.

\bigskip

{\noindent {\em Note added.}}\hspace{2mm} While finishing this paper,
two related papers \cite{rDVV,rNeumann}\ appeared on the hep-th
archive.

\end{document}